\documentclass[12pt]{article}
\usepackage{amsmath}
\usepackage{amssymb}
\usepackage{epsfig}
\usepackage{color,colordvi}
\def\colour4colour#1{\Blue{#1}}

\newcommand{\eq}{\begin{equation}}
\newcommand{\eqx}{\end{equation}}
\newcommand{\eqn}{\begin{eqnarray}}
\newcommand{\bi}{\begin{itemize}}
\newcommand{\eqnx}{\end{eqnarray}}
\newcommand{\ei}{\end{itemize}}

\newcounter{hran}

\def\MSbar{\relax\ifmmode\overline{\rm MS}\else{$\overline{\rm MS}${ }}\fi}

\renewcommand{\baselinestretch}{1.5}

\begin{document}
\begin{titlepage}
\begin{center}

\hfill CERN-PH-TH/2009-125\\
\vskip -.1 cm
\hfill IFT-UAM/CSIC-09-XX\\
\vskip -.1 cm
\hfill MPP-2009-130\\

\hfill arXiv:yymm.nnnn\\

\vskip .2in

{\LARGE Strong Coupling Holography}
\vskip 0.1in

{\bf Gia Dvali$^{1,2,3}$ and Cesar Gomez$^{4}$}

\vskip 0.1in

${}^1\!$
Theory Division, CERN\\
CH-1211 Geneva 23, Switzerland\\
%{\footnotesize{\tt }}
%\vspace{.1in}

{\sl $^2$  CCPP, Department of Physics, New York University\\
4 Washington Place, New York, NY 10003 }
%\vspace{.1in}

{\sl $^3$  Max-Planck-Institute for Physics,\\
F\"ohringer  Ring 6, D-80805, M\"unchen,  Germany}

%\vskip .2in

${}^4\!$
Instituto de F\'{\i}sica Te\'orica UAM-CSIC, C-XVI \\
Universidad Aut\'onoma de Madrid,
Cantoblanco, 28049 Madrid, Spain\\
{\footnotesize{\tt georgi.dvali@cern.ch , cesar.gomez@uam.es}}

\end{center}

\vskip .1in

\centerline{\bf Abstract}
\vskip .1in
%\no

\noindent
 We show that whenever a 4-dimensional theory with $N$ particle species  emerges as a consistent low energy description of  a  3-brane   embedded in an asymptotically-flat 
 $(4+d)$-dimensional space,  the holographic  scale of high-dimensional gravity sets the strong coupling scale of the 4D theory. 
%  The memory about the high-dimensional holographic scale persists in form of the  strong coupling scale of the 4D theory  
  This connection persists in the limit in which gravity can be consistently decoupled. 
   We demonstrate this effect  for  orbifold planes,   as well as for the solitonic  branes  and string theoretic 
   $D$-branes.   In all cases the emergence of a 4D strong coupling scale from bulk holography 
 is a persistent phenomenon.  The  effect turns out to be insensitive even to such extreme deformations of the 
 brane action that seemingly  shield  4D theory from the bulk gravity effects. A well understood example of such deformation  is given by large 4D Einstein term in the 3-brane action, which is known to 
 suppress the strength of 5D gravity at short distances and change the  5D Newton's law into the four-dimensional one.     
% create an effectively-4D weak gravity at short distances.  
 %Due to this weakening of gravity, the resolution length-scale of  $N$ 4D-species is decreased  well beyond the high-dimensional holographic bound, seemingly avoiding the latter.  
 Nevertheless,  we  observe that the scale at which the scalar polarization of an effective 4D-graviton becomes  strongly coupled is again set by the bulk holographic scale.  The effect persist in the gravity decoupling limit, when the full theory reduces to a 4D system  in which the only memory about the high-dimensional holography is encoded in the strong coupling scale. 
 The observed intrinsic connection between the high-dimensional flat space holography  and 4D strong coupling  suggests  a possible guideline for  generalization of AdS/CFT  duality to other systems.       

\vskip .4in
\noindent

\end{titlepage}
\vfill
\eject

\def\baselinestretch{1.2}

%%%%%%%%%%%%%%%%%%%%%%%%%%%%%%%%%%%%%%%%%%%%%%%%%%%%%%%%%%%%%%%%%%%%%%%%

\baselineskip 20pt

%%%%%%%%%%%%%%%%%%%%%%%%%%%%%%%%%%%%%%%%%%%%%%%%%%%%%%%%%%%%%%%%%%%%%%%%
%%%%%%%%%%%%%%%%%%%%%%%%%%%%%%%%%%%%%%%%%%%%%%%%%%%%%%%%%%%%%%%%%%%%%%%%

\section{Introduction}
  An extraordinary example of  connection between gravity 
 and quantum fields is  AdS/CFT correspondence \cite{ADS}. According to  it the 
 quantum dynamics of fields can be understood in terms of classical gravity in higher dimensions
 and vice-versa. 
 
   In search of  the  gravity/field theory duals for the other  systems, one has to be prepared that 
   the power of extended duality may be more limited than in the AdS/CFT case.  The new  dualities may 
   be more useful in  relating certain quantities and less useful for  others. 
  Nevertheless,  even a limited  generalization  of AdS/CFT correspondence  would be of fundamental 
  importance.     
  
   In trying to discover  such generalizations, it is important to identify  relations that are not 
   unique to the AdS geometry of the setup.  
  In the present paper we shall identify such a relation, which,  although shared by AdS/CFT,   is not 
  bounded to the existence of AdS geometry in a near horizon limit. 
  
    We shall observe,  that an emerging holography \cite{tHooft}  will play the crucial role in our findings.  
  
  The  key quantities of our focus will be,  the number of species on 4D field theory side and 
  the holographic scale for $N$ bits of information on the high-dimensional gravity side.   Our central  finding is,  that the strong coupling scale of  such 4D theory is determined by the bulk holographic 
  scale and this property extends far beyond the AdS/CFT setup.  
  
    Before formulating our results quantitatively,  we shall first describe  the  framework we shall be working in.   We start with setting the connection between the  number of species and the various  scales of the theory.  
 
    It has been shown by recent studies \cite{bound, us}, that  consistency of the well-understood black hole (BH) physics demands that in any $4+d$-dimensional  Einsteinian theory of gravity  with  $N$ propagating particle species  the upper bound on the gravitational cutoff of the theory is given by  the following energy scale, 
 \begin{equation} 
\label{cutoff}
  M_{N(4+d)} \, \equiv \, {M_{4+d} \over N^{1\over 2+d} } \,  ,  
\end{equation}
where $M_{4+d}$ is the $4+d$-dimensional Planck mass.  The corresponding Planck length we shall denote  by $L_{4+d}$.   Eq(\ref{cutoff})  follows from the fact,  that 
the BHs smaller than the length scale 
  \begin{equation}
 \label{lbound}
 L_{N(4+d)} \, \equiv \, N^{{1\over 2+d}} \, L_{4+d} \, ,  
\end{equation}
inevitably violate quasi-classicality conditions (such as,  the heating rate not exceeding the 
temperature squared)  and can no longer be treated as such (see \cite{bound, us} for details).  

 The hierarchy (\ref{cutoff}) between the Planck mass and the cutoff
is  expected due to renormalization of the graviton kinetic term by species loops \cite{loops1,loops2}. However the power of the BH argument is that it is fully non-perturbative and directly applies to the physical values, regardless of perturbation theory artifacts, such as the possible cancellations or fine tunings of the loop contributions.   

  Before proceeding we wish to comment on some other  bounds on  species number 
 and stress  physical differences from our case.  The black hole quasi-classicality scale
  was used in \cite{Br1}, but  the bound on the cutoff obtained in this work is milder,  $M_4/{N}^{1/4}$,  as opposed to $M_4/\sqrt{N}$ in our case. \footnote{Interestingly, this scale coincides with a three-level unitarity bound on production  of species in one-graviton exchange  amplitude  (see the fifth  reference in \cite{bound}), although the authors of \cite{Br1} arrived there by other means.}  The scale $M_4/\sqrt{N}$ was first obtained as the upper bound on the temperature of radiation-dominated universe in which $N$ species are in thermal equilibrium in \cite{Bek1} (and further generalized in \cite{Br2}). 
 The crucial difference in the meaning of this scale in our case \cite{bound,us}  is,  that it is an absolute 
 cutoff of the species theory, at which gravity {\it of individual} species  gets strong and the species resolution is no longer possible.  Therefore, for example, one 
 cannot raise the temperature of the system above this scale (while staying in a weak gravity regime) even if only a single flavor of species is in equilibrium and all the remaining $N-1$ species  are decoupled. 
 
   To summarize, the scale (\ref{cutoff})  in our case is the scale above  which  gravitational interactions of elementary particles cannot continue to be in a weakly-coupled classical regime. 
   
%    Instead,  it is  an  absolute upper bound on the weak gravity scale in any theory with $N$ species.    

 A particular  usefulness of the relation (\ref{cutoff}) is that  it enables understanding of  geometric  relations  in terms of species number counting.    A simple example of such counting  is given by 
Kaluza-Klein  (KK)  theory of $(D + d)$-dimensional space in which $d$-dimensions 
are compact, and the metric on non-compact $D$-dimensions is Minkowskian.   The relation between the $D$-dimensional  and $(D + d)$-dimensional  Planck scales is given by, 
 \begin{equation}
 \label{mplancks}
  M_{D}^2 \, = \, M_{D+d}^2  \,   V_{(d)} ,  
\end{equation}
where $ V_{(d)} $ is the volume of $d$-dimensional  compact  space measured in the units of  the fundamental  Planck length $L_{D+d}$.   In each concrete case finding  $V_{(d)}$ can be a complicated 
geometric task, but at the end of the day it is given by the number $N$ of KK  species as counted by 
a $D$-dimensional observer.     
  
   We have shown recently \cite{us},   that the dependence (\ref{cutoff})  of the gravitational cutoff of $4+d$-dimensional theory on the number  $N$ ($4+d$-dimensional) particle species, 
%\begin{equation} 
%\label{cutoff}
%  M_{N(4+d)} \, \equiv \, N^{-{1\over 2+d}} \, M_{4+d} \, ,  
%\end{equation}
can be easily understood from the quantum information point of view.   Indeed, having $N$ particle species means that we can distinguish and label them by physical measurements. 
That is, we can encode information in the species label.  The critical length scale 
below which we can no longer read the species label is inevitably a cutoff of the theory.  
The role of such critical length is precisely played  by (\ref{lbound}).
Indeed,  the distance $L_{N(4+d)} \, \equiv \, 1/ M_{N(4+d)}$ marks the size of a smallest pixel that can decode the species label.  Such a pixel has to contain samples of all the $N$ species in order 
to identify the label of any incoming particle. 
 Since localization of a particle wave-function in a box of size $L_{pixel} $ cost at least energy 
 $1/L_{pixel}$, the lowest mass of the pixel is therefore 
$m_{pixel} \, = \, N/L_{pixel} $.  The bound on $L_{pixel}$ is imposed due to requirement that the pixel does not collapse into a BH.  In other words, $L_{pixel}$ has to exceed the corresponding Schwarzschild radius of the pixel ($r_{pixel}$), which in 
$4+d$ dimensional gravity is given by 
\begin{equation}
\label{radius}
 r_{pixel} \, = \, \left ({N \over L_{pixel}}  {1 \over \, M_{4+d} ^{2+d} } \right ) ^{{1 \over d + 1}}\, .
 \end{equation} 
Requiring $r_{pixel} \, \leqslant   \, L_{pixel} $,  we get  the bound
 \begin{equation}
 \label{lscale}
  L_{pixel}  \, \geqslant  \, L_{N(4+d)} \, .  
\end{equation}

Notice, that the scale  $L_{N(4+d)}$ is at the same time the holographic bound for $N$ information bits. 
This is natural, since the species label is a particular form of information, and its storage must obey 
the general laws of holography. 
 
  In the present paper we shall be interested in the situation in which the $N$ species live in a 4D sub-space and do not propagate in the whole $4+d$ space-time.   Such a situation is physically motivated, 
  since localized zero modes are generic in theories with branes.   We thus  consider a situation in 
  which bulk theory is $4+d$-dimensional gravity (with $\sim 1$ species) propagating on 
  asymptotically-flat background and  interacting with a  4D 
  theory with $N$ particle species.  We shall investigate how the 4D species cope with the laws of bulk holography. 
  
   We shall discover that the bulk holographic scale (\ref{lbound})  plays a crucial role in determining the strong coupling scale of the 4D theory ($\Lambda_{str}$).  Namely,  the following bound holds,  
  \begin{equation}
 \label{strongbound}
  L_{N(4+d)}   \, \leqslant  \, L_{str} \, ,   
\end{equation}
 where $L_{str} \,  \equiv \, \Lambda_{str}^{-1}$.     
 
     This role is robust  and persists even  for  extreme deformations of the 
   4D action, that seemingly shields the brane theory from the bulk gravitational effects. 
   Even in the limit when one consistently decouples gravity, the system reduces to a  pure 4D theory 
   in which the strong coupling scale is set by the bulk holographic bound.
   
    The connection to AdS/CFT is explicit, since as we shall see, the latter is an example of saturation of the above bound. 
 In AdS/CFT  the role of our species is played  by open string zero modes living on the stuck 
 of $D_3$-branes embedded in asymptotically-flat  10D space.  The  strong coupling scale of 
 4D theory is set by the 10D holographic scale.  This property turns out not to depend on the AdS nature 
 of the  near horizon geometry, which otherwise is crucial for establishing AdS/CFT.
  The same effect is exhibited by other cases in which species are confined to the branes that do not 
  exhibit AdS properties in near-horizon limit.      
Our observations suggest a possible  high-dimensional holographic  origin  of  the strong coupling scale
of 4D theories,  which  extends beyond the AdS/CFT correspondence.

\section{Holography for Localized Species}

Consider a  5D  space-time with coordinates $x_{\mu},  y$ ($\mu  = 0,1,2,3$),    in which $N$  species are localized on a 3-dimensional  surface (a  $3$-brane), located at  $y \, = \, 0$.   We shall assume that the bulk space-time metric, $g_{A,B}(x_{\mu}, y) $, is asymptotically flat, 
and the bulk theory is a pure 5D  Einsteinian gravity  with 5D Planck mass $M_5$. 
 Since below $M_5$ the only species propagating in the bulk is the five dimensional graviton,  
  the natural cutoff of the bulk theory  is 
$M_5$. The corresponding fundamental length scale we shall denote by $L_5 \, \equiv \, M_5^{-1}$. 
  In particular, $L_5$ is a lower bound on the size of quasi-classical BHs in the bulk.
   What is the fundamental scale for the brane theory? 
  
  The brane theory is a 4D theory of $N$ species coupled to 5D gravity.  The fundamental holographic scale of this theory is set by the shortest length-scale on which the resolution of species is in principle 
  possible.  That is,  cutoff is set by the size of a smallest pixel in which one is able to  localize $N$ species.  The requirement that such a pixel does not collapse into a BH under the influence of 5D gravity, fixes the lower bound 
 \begin{equation}
   L_{N5}  \, = \,  N^{1/3}\, M_5^{-1} \, ,
   \label{l5} 
  \end{equation}
 which coincides with the holographic scale of the 5D theory for $N$ bits of information.   
  This coincidence is no accident, since as we have explained,  the species flavor is a particular form of the information, and has to obey the same holographic bounds.    
   
   Notice,  that the same scale represents  the lower bound for the BH quasi-classicality, since any 
  5D  BH of size $R < L_{N5}$  intersecting with  the brane, would half-evaporate during a time less than its size, which is a clear violation of quasi-classicality conditions.  
   
    We are thus lead to the following  crucial conclusion: 
    
    {\it For a $3$-brane theory with $N$ species, the lower bound 
    on the cutoff scale is set  by the holographic bound of  5D bulk theory for  $N$ bits of information storage. } 
    
    In the other words beyond the scale  $L_{N5}$ some change of the regime must occur in the
    4D theory.  What is this change of the regime? 
  
   There are the following two logical possibilities.
    
   First,  let us assume that laws of gravity are  essentially unaffected by the presence of the brane, meaning that   4D sources continue interacting via  5D Einsteinian gravity  for the distances 
    $r \, \ll \, L_{N5}$.   In such a case, we are forced to admit that  the species can no longer be emitted 
    as elementary states from the BHs of intermediate size $L_5 \, \ll  \, r_g \, \ll L_{N5}$, because if they were, this would be in conflict 
    with the assumption of 5D BH quasi-classicality all the way till $L_5$.          
   Thus, the identities of  species must be 
  compromised  at shorter distances,  and the species theory must be UV completed. 
  For example,  species may  simply become composite  below the scale  $L_{N5}$.
  
    On the other hand, if we assume that  species continue to be elementary for distances 
  $\ll  L_{N5}$, then we are inevitably lead to the conclusion that  5D gravity  must change the regime beyond this scale.        
    
   To summarize,  the  fact that in the presence of the localized species the  5D theory breaks down at distances  $L_{N5}$,  suggests that 
 either  effective  4D  field theory of  $N$ species  or gravity  (or both) require an UV completion beyond  
 the bulk holographic scale.   
 
  Insisting that species remain elementary below $L_{N5}$ leaves us with inevitability of 
  the gravity-modification at  some crossover distance  $r_c \,  \geqslant \, L_{N5}$.    
    We shall now discuss the underlying physics of this modification.
For transparency of presentation, where possible, all unessential order-one  numeric factors will be absorbed in definition of scales. The reader can easily reproduce them, whenever needed.

  \section{Gravity of Species on a Tensionless Brane (Localization on Orbifolds)} 
  
    We shall consider first the case of a tensionless brane.  In such a case, the brane is effectively a 
    surface at the  $Z_2$-orbifold fixed point.  
     
  The curious fact is that, despite the zero tension, such a brane continues to  have dramatic  gravitational effect.  
 Indeed, as explained above,  the classical  5D gravitational description is impossible to be valid at distances shorter  than $L_{N5}$, without compromising the validity of $N$-species description above this scale. 
  If we insist on the validity of 5D description all the way till $L_{5}$, this will lead us to the conclusion that the bulk and the brane observers must experience the different cutoffs.  
 The cutoff of the bulk theory remains  $M_5$, whereas the cutoff of the world-volume  4D theory 
is  $L^{-1}_{N5} \, \ll \, M_5$, since as explained above,  the  brane observer cannot resolve species in her own theory above this energy scale.     
    
    What if we impose the requirement that the cutoffs of the two observers  must be equal? 
   This means, that  the brane observer should be able  to continue resolving species down to the distances  $L_5 \,  \ll \, L_{N5}$. This is only  possible if  the brane gravity starts becoming {\it weaker} than the  5D Einsteinian gravity  at distances smaller than a certain critical size $r_c$,  
 which  must be larger than the 5D holographic scale,   $r_c \,  >  \,  L_{N5}$. 
 
  Indeed,  if the brane observer were to distinguish species down to the distances $L_5$,  
    the  brane gravity must be such that the Schwarzschild radius of a BH of mass $N/ L_5$, must not exceed  $L_5$.        
  Thus,  below some distance scale $r_c \, \gg \, L_{N5}$, the brane gravity should cross over to a regime in which the  gravitational  radius of a given  source is shorter than in 5D Einstein gravity.  
   That is, for  distances $r < r_c$ the Newtonian potential must become  
   \begin{equation}
   V(r < r_c) \, = \, {1 \over M_5^3 r_c^{\alpha} }  {1 \over r^{2- \alpha}} \, , 
  \label{newgravity}
  \end{equation}
  where $\alpha \, > \,  0  $ is a parameter  to be determined in a moment.
 The crossover scale $r_c$ is fixed from the condition that under the gravitational law (\ref{newgravity}) the gravitational radius of  a pixel 
 of size $L_5$ with $N$ localized species must be equal to $L_{5}$. That is, 
 \begin{equation}
 r_c \, = \, N^{{1\over \alpha }} \, L_5  \, ,
 \label{rc}
 \end{equation}
 
  We shall now see that $\alpha \, = \,  1$, because of the following reason. 
  Modification of gravity ($\alpha \neq 0$) implies that the brane has a non-trivial gravitational effect
  on the probe sources.  
  This effect can only be manifested through the terms in the brane's gravitational action. 
  Since the brane tension was by construction set to zero, the only remaining relevant term, that respects the world-volume  general covariance,  is the 4D Einstein-Hilbert  term of the induced metric
  \begin{equation}
  \int \, d^4x \, \sqrt{-g}\, M_{4}^2  \, R_{4} \, ,  
  \label{r4}
  \end{equation}
  where $g_{\mu\nu}(x) \, \equiv \, g_{\mu\nu} (x, y =0)$ is the induced metric on the brane, and we have labeled the coefficient by $M_4^2$.   
  The gravitational effect of this term is well-studied \cite{dgp}, and it is known to lead to exactly the above-discussed  modification  of gravity at the crossover scale $r_c = M_4^2 /M_5^3$  and  with $\alpha = 1$!
   The brane-brane graviton propagator  takes the following form  (tensorial structure, which is the one of massive spin-2, is suppressed)   
   \begin{equation}
 G(p) \, = \, {1 \over M_5^3r_c} {1 \over p^2 \, + \, \sqrt{p^2}/r_c} \, ,   
   \label{propagator}
   \end{equation}    
 where $p$ is the usual four-momentum. 
 This propagator  reproduces exactly the above cross-over behavior. 
  At distances $r \, \ll \, r_c$ gravity crosses 
  over to a 4D regime with the effective 4D Planck scale given by
\begin{equation}  
  M_4^2 \, = \, M_5^3r_c \, = \, N M_5^2 \,. 
\end{equation}

  Notice,  that this is exactly the value that is expected from the 
  contribution of the loops of localized $N$ particles \cite{loops1, loops2, loops3}.  The important input from  the holography 
  is, that the crossover behavior has to be there because of fully non-perturbative consistency 
  reasons, irrespective of the loop contribution. 
  
   The physics of the above crossover behavior is very transparent  in terms  of 
   KK-decomposition of the 
   5D graviton.  On any 4D-Lorentz-invariant  background such a decomposition has the following general form, 
   \begin{equation}
   h_{\mu\nu} (x, y) \, = \, \int_0^{\infty}  \, dm \,  \psi^{(m)}(y)
\,  h^{(m)}_{\mu\nu}(x) \, ,
   \label{kkexpansion}  
   \end{equation} 
   where  $h^{(m)}_{\mu\nu}(x)$ are the 4D massive spin-2 fields and  $\psi^{(m)}(y)$  are their wave
   function profiles in $y$-dimension that form a complete orthonormal set. 
    The effective 4D graviton,  to which a brane observer couples, is a collective mode  given by 
    the superposition of infinite number of KK states,
   \begin{equation}
   h_{\mu\nu} (x) \, \equiv  \, h_{\mu\nu} (x, y =0) \, = \, \int_0^{\infty} \, dm \,  \psi^{(m)}(0) \, h^{(m)}_{\mu\nu}(x) \,. 
   \label{resonance}  
   \end{equation} 
  The propagators of $h_{\mu\nu}^{(m)} (x)$-states are the usual 4D propagators of the 
  massive Pauli-Fierz fields, which mediate Yukawa type potentials,   ${{\rm e}^{-mr} \over r}$.  
  The non-trivial information about the changes of the gravitational regime is encoded in the 
  boundary values of $\psi^{(m)}(y)$ functions at $y=0$.  The key point is that  because of $R_4$-term on the brane, the  wave-functions of $m \gg r_c^{-1}$ modes are suppressed at $y=0$, 
  \begin{equation}
  |\psi(0)^{(m)}|^2  \, = \, {1 \over 1 \, + \, (r_cm)^2} \,,
  \label{psi}
  \end{equation}
(for simplicity we have set numeric factors equal to one).  
 As a result of this suppression, the brane observer is effectively decoupled from the KK modes 
 that are heavier than $r_c^{-1}$, and sees the following effective gravitational potential 
  \begin{equation}
 V(r)_{brane} \, = \, {1 \over M_5^3} \,   \int_0^{\infty}  \, dm \,  |\psi(0)^{(m)}|^2  \,  {{\rm e}^{-mr} \over r}\, \simeq \,
 {1 \over M_5^3}\, \int_0^{r_c^{-1}}  \, dm  \,  {{\rm e}^{-mr} \over r}\,  \, 
  \label{branepot}
  \end{equation} 
where in the last term we had to cutoff the integral at $r_c^{-1}$ by taking into the account the suppression 
of the heavier modes given by (\ref{psi}).    Due to this suppression,  at the distances $r \ll r_c$ the brane observer sees a weak 
4D gravity 
 \begin{equation}
 V_{brane}(r \ll r_c)  \, \simeq \, {1 \over M_5^3r_c} {1 \over r}\, .
  \label{branepot1}
  \end{equation} 
This goes in sharp contrast with the experience of the  bulk observer locates say at $y\, \gg \, r_c$. 
 Such an observer couples to an entire KK tower and 
experiences 5D gravity for all $r$,   
  \begin{equation}
 V(r)_{bulk} \, = \,  {1 \over M_5^3} {1 \over r^2} \, .
  \label{bulk}
  \end{equation} 
    
   Thus,  because of the modified brane gravitational action, imposed by holography,    
the  brane observer at short distances sees 4D gravity that is much weaker than its would-be 5D 
counterpart.  The crucial fact is,  that  the crossover scale $r_c \, = \, N \, L_5$ at which the 4D regime sets in is much larger than 
 the 5D holographic scale $L_{N5}\, = \, N^{1/3}$, resulting into the following hierarchy of scales 
   \begin{equation}
   (r_c \, = \, N \, L_5) ~  \gg  ~ (L_{N5} \, = \, N^{{1\over 3}} L_5) ~ \gg ~  L_5 \, .
   \label{hierarchy}
   \end{equation}
 In the other words, the brane observer manages to store the $N$ bits of information 
 in a box of size $L_{pixel} \, = \, L_5$,  which is much smaller than the size of the analogous box ($L_{N5}$) required for the storage of exactly the same amount of information  by  her 5D counterpart.  This is possible  
 thanks to stretching the crossover distance $r_c$  much beyond the 5D holographic scale  $L_{5N}$.  
 
    The above situation  creates an impression that thanks to crossover modification of gravity the 
  4D-observer managed to resolve species  all the way  to distances $L_{pixel} \, = \, L_5$, and in this way bypass 
  the constraints of 5D holography. 
   We are now ready to ask, whether  because of  modification of gravity  the memory about the 5D holographic cutoff  is completely erased  in 4D theory.   Remarkably, the answer to this question is 
   negative.  
 Although crossover modification of gravitational regime  extends
 the  species-resolution scale  to  the distances much shorter  than $L_{N5}$, nevertheless,  the latter scale still manifests itself  through  the strong coupling scale in the 4D theory! 
  The key for establishing this connection is the appearance of a new intermediate strong coupling scale  in 4D theory, which we shall now consider.

\subsection{Strong coupling scale of scalar  gravitons and $r_*$-phenomenon}
  
   By now It is very well understood \cite{strong, prl, stronggia} that  in the above model the scalar polarization   
of an effective  4D graviton (\ref{resonance}) has self-interactions that are suppressed 
by the new scale,   
\begin{equation}
\label{strongscale}
\Lambda_{str} \, \equiv  (r_c^{-2}\, M_P)^{{1\over 3}} \, . 
\end{equation}
The origin of the latter strong coupling scale is the following. 
As we have seen,  the 4D graviton (\ref{resonance}) that is sourced by  4-dimensional 
energy-momentum tensors living in the brane world-volume theory is 
not a mass-eigenstate, but  a resonance representing a linear  superposition of the  continuum of the massive KK modes. 
Since each KK graviton is a massive spin-2 boson, it contains five physical degrees of freedom, 
 and so does their  composite  4D graviton  (\ref{resonance}).  In fact,  from the 4D perspective the decomposition (\ref{branepot}) is nothing but a spectral representation of the resonance, 
 with the spectral function given by (\ref{psi}).    
 
 One of the five physical degrees of freedom  is an extra scalar polarization, which we shall denote by 
$\pi$.  Ignoring the spin-1 helicity, in terms of canonically normalized spin-2 
Einsteinian helicity ($\tilde{h}_{\mu\nu}$) and the 
scalar helicity ($\pi$), the 4D graviton (\ref{resonance}) can be decomposed as, 
  \begin{equation}
\label{handpi}
h_{\mu\nu} \, = \, \tilde{h}_{\mu\nu} \,  + \, {1\over 6} \eta_{\mu\nu}\pi \, + \,
 {r_c \over 3}  {\partial_\mu \partial_\nu  \over \sqrt{\Box}}\pi\,.
\end{equation}
Straightforward  analysis \cite{strong} shows,  that
 while  the Einsteinian  spin-2 helicity $\tilde{h}_{\mu\nu}$ experiences  the usual $1/M_4$-suppressed interactions,  the scalar mode $\pi$ becomes self-strongly coupled at much lower scale  $\Lambda_{str}$, given by (\ref{strongscale}). 

 The effect of this strong coupling on the  gravitational field created by localized sources  is rather  profound and leads to creation of a new intermediate 
 gravitational scale \cite{strong, stronggia, andrei, lue,  moon, ls, greg, oriol, nicol, ghostcedric}, 
  \begin{equation}
\label{rstar}
r_* \, \equiv \, \left(r_c^2r_g\right)^{{1 \over 3}} \; .
\end{equation} 
 Obviously, for any source with the Schwarzschild radius $r_g \, \ll \, r_c$, we have the following hierarchy of scales,      
\begin{equation} 
 \label{between}
 r_g \, \leqslant \,  r_*  \,  \leqslant \,  r_c  \, . 
 \end{equation}
 
  The notion of $r_*$ for the above 5D theory \cite{dgp}  is analogous to the so-called Vainshtein
effect \cite{arkady} that avoids van Dam -Veltman-Zakharov discontinuity \cite{vDVZ}  in non-linear theory of Pauli-Fierz massive graviton. \footnote{However, 
as shown in \cite{ghostcedric, ghostnicol, degrav}, for non-linear Pauli-Fierz case the Vainshtein effect is  intertwined with the appearance of  Boulware-Deser type \cite{BD} ghost, which is absent in the present case.}

 The physical meaning of the  scale $r_*$ can be defined as the distance from the source at which the non-linear interactions of the scalar  helicity $\pi$  become important.  In this sense,  the radius $r_*$ plays the same role for $\pi$  as the ordinary Schwarzschild radius  
 $r_g$ plays for Einsteinian  spin-2 helicity $\tilde{h}_{\mu\nu}$.   The existence of $r_*$  was  demonstrated by explicit  solutions\cite{strong, andrei, lue,  moon, ls, greg, oriol, nicol, ghostcedric}. However,  emergence  of this scale  can be understood already from the simple power counting arguments \cite{strong, stronggia}.  
 
 Consider a 4-dimensional static source of the Schwarzschild radius  $r_g$ and imagine that we are trying to probe its metric.  At sufficiently large distances $r \gg r_*$  (low momenta  $p \ll r_*^{-1}$)  the interaction between the source and the probe 
 are dominated by one-graviton exchange.   At some shorter distances (high momenta)  non-linear self-interactions 
 of gravitons become important. This critical distances will define the value of $r_*$.   
The leading non-linear coupling comes  from the trilinear 
interaction of the longitudinal gravitons ($\pi$) with the coefficient  $r_c^{2}/M_4$, and  the corresponding  vertex has a momentum dependence of the form
\begin{equation}
\label{vertex}
      {r_c^{2}\over M_4} \,p^4 \,,
\end{equation}
where $p$ is a typical energy-momentum flowing through the vertex.
The  scale $r_*$ corresponds to the distance from the source at which the contribution 
from the above trilinear vertex becomes as important as the linear contribution given by one graviton exchange. This fixes (\ref{rstar}).
%
% \begin{equation}
%\label{rstar}
%r_* \, = \, \left(r_c^2r_g\right)^{{1 \over 3}} \; .
%\end{equation} 

In order to single out the leading non-linearity,  we can also take an useful limit  $r_c, M_4 \,  \rightarrow \,  \infty$ and  keeping $\Lambda_{str}$ fixed \cite{prl, degrav}. In the language of species, this corresponds to taking $N, M_5 \, \rightarrow \, \infty$.  In this limit, 5D gravity decouples. However, the resulting theory is non-trivial. It is  
a 4D theory of a single scalar field $\pi$ with the 
Lagrangian
 \begin{equation}
\label{pidecoupling}
\pi \square \pi \, 
 +  \,  {1 \over \Lambda_{str}^{3} } (\partial_{\nu} \pi)^2   \square \pi \, \, +  \ { \pi \over  M_4} \, T \, , 
\end{equation}
where the strong coupling scale $\Lambda_{str}$ is  given by (\ref{strongscale}), and $T \, \equiv \, T_{\mu}^{\mu}$ is an external source.  As argued in \cite{greg2}, this form may not capture some  essential properties of 5D theory, but this difference is inessential for our purposes, since as explained above,  the strong coupling effect is well established already  in the full theory.  

Notice that in order to maintain a non-trivial source in the decoupling limit, we should 
keep  $T/M_4$ fixed.  This means that in this limit  $r_g \, = \, T/M_4^2 \, \rightarrow \, 0$, but $r_*$ remains fixed.  The above-explained physical meaning of $r_*$ can be made explicit by exactly solving the 
equation for $\pi$, which is effectively an algebraic equation on the derivative of $\pi$\cite{nicol}.  We shall not do this,  since it is enough to notice that at sufficiently  large distances (where the non-linear term is unimportant) 
the solution must be the usual  $\pi \, = \, r_g /r$.  Then, comparing  the strengths of the variations of the  two terms in 
(\ref{pidecoupling}) on this function, it is easy to see  that the linearized solution breaks down and gets modified exactly at the scale $r_*$, where the non-linear  term in the equation of motion  becomes dominant.  We shall now see that the scale $r_*$  is the holographic bridge between the 4D  and 5D theories.

\subsection{Holographic Meaning of $r_*$}

 We are now ready to discuss how the bulk holographic scale manifests itself in an effective 4D theory. 
The fundamental length scale of 5D theory is $L_5$.  This scale represents an absolute lower bound 
on the localization length of a single information bit, since it is a critical distance at which  
the size of the bit ($L_{bit}$)  and its  5D  gravitational radius ($r_{bit}^{(5)}$) cross. 
 In other words,  the Schwarzschild radius 
of any localized bit of length $L_{bit} = L_5$  is the same as the length itself.    

 The energy of such an  information bit  is $m_{bit} \, = \, M_5$, whereas the 5D  gravitational radius is $r_{bit}^{(5)} \,    = \,  L_5$.  Since in 5D the Schwarzschild radius scales as the cubic root of the mass, 
 the holographic bound on localization of  $N$ information bits
 is $L_{N5} \, = \, N^{{1\over 3}} L_5$.   

  Now, as discussed above,  by populating the 4D world-volume theory by  $N$ particle species and requiring  that species identities stay resolved all the way to distance $L_5$, we are forced to 
the conclusion that at short distances gravity on the brane  must be weaker  than 5D gravity. This implies the existence of 
4D Einstein-Hilbert term (\ref{r4})  with the coefficient  $M_4^2 \, = \, N \, M_5^2$. 
  
  Thus, from the first glance it seems that the 4D observer 
  has  avoided the 5D-holographic  constraint on the information storage.  Indeed,  the mass of an elementary information  bit  in 4D and 5D theories is the same,  simply given by the inverse localization 
 length,   $m_{bit} \, = 1/L_{bit} \, = \, M_5$.  As explained above,  in 5D gravity  the Schwarzschild radius of such information  quantum  would be $r_{bit}^{(5)} \, = \, L_5$. 
 However,  in 4D brane theory the Schwarzschil radius of the same bit  ($r_{bit}^{(4)}$)  is now set by the weaker 4D gravity,  
 \begin{equation} 
 r_{bit}^{(4)} \, = \,  {M_5 \over M_4^2} \, = \, {r_{bit}^{(5)}  \over N} \, ,
 \label{rbit4}
 \end{equation}
 and correspondingly is much shorter  than its 5D counterpart.   This  creates an impression that 
 4D theory is completely liberated from the 5D holographic bound, and has no memory about $L_{N5}$. 
 
   Remarkably, this is just an illusion, and  the 4D system still keeps a clear memory about the 5D holography.   This memory manifests itself through the fact that the   $r_*$ radius 
 of the above elementary information bit,  with the Schwarzschild radius $r_{bit}^{(4)}$ given by (\ref{rbit4}),  is   
 \begin{equation}
 r_* \, = \,  (r_{bit}^{(4)} \, r_c^2)^{{1\over 3}} \, = \, N^{{1 \over 3}} \, L_5 \, .  
 \label{rstar4}
 \end{equation}
 This is exactly equal to the 5D holographic scale $L_{N5}$!
  
   More generally,  irrespective  of an origin or a value of  $r_c$,  the $r_*$ radius  of $N$ information bits in 4D theory  scales with $N$ exactly as the 5D holographic scale,  
\begin{equation}
 \label{scaling}  
  r_* (N) \, \propto \,  L_{N5} \, \propto\,   N^{{1\over 3}} \, . 
 \end{equation} 
 
%   Let us define $N_{eff}$ as the number of  bits (species) that being localized in a box 
 %of size $r_c$ would cost the same energy.  In the other words,  
 %\begin{equation}
 %\label{neff}
% E_N \, = \, {N_{eff} \over r_c}  \, = \, {N \over L_{N5}} \, . 
%\end{equation} 
%Computing now the $r_*$ radius of  the above mass $E_N$, we get, 
%\begin{equation}
%\label{nbits}
%r_*(N) \, = \, \left ({N \over L_{N5} M_5^3r_c} r_c^2 \right )^{{1\over 3}} \, = \, {N_{eff}^{{1 \over 3}} \over M_5} \, 
%\end{equation}
%which is exactly the 5D holographic bound for $N_{eff}$ information bits! 

Thus,  what we are finding is,  that 5D holographic scale secretly penetrated in 4D theory as a scale 
of strong interactions.

   The generalization of the above  holographic counting to higher co-dimension  tensionless rigid branes is straightforward.   For  a $3$-brane with localized $N$ species embedded in $4+d$-dimensional space,  the  holographic scale that sets  the 4D cutoff is given by (\ref{lbound}).   
  % \begin{equation}
   %L_{N(4+d)} \, = \, N^{{1 \over 2+d}} L_{4+d} 
  %\label{ld}
   %\end{equation}
  %where,  $L_{4+d}$  is the $4+d$-dimensional Planck length.
   The  BH quasi-classicality arguments tell us that  resolution of species identities 
  at distances shorter than $ L_{N(4+d)} $ is incompatible with 5D gravitational regime 
  beyond  this scale.  And again, should we demand that 4D observer  be able to resolve species 
  all the way down to the $(4+d)$ Planck length, we shall be confronted with the necessity of the crossover modification of gravity at some scale $r_c \, \gg \, L_{N(4+d)}$,  with the subsequent strong coupling 
  effect at the short distances.

  %    Thus, our generalized key point  is,  that  for a $3$-brane theory with $N$ localized species, 
   %the lower bound on the scale of gravity modification is set by the holographic bound of the  bulk 
   %$4+d$-dimensional gravity theory for  $N$ bits of information storage.  
   %The above bound is model independent.  

% However,  if we require the equality of the two observer's  cutoffs,
% again we will be lead to the conclusion that gravity has to cross over to a four-dimensional regime 
 %below the distance $r_c \, \gg \, L_{N(4+d)}$. This fact also remains unaltered.   
    %However,  for higher co-dimension cases the scaling of   $r_c$ with $N$ is  more subtle then in the co-dimension-one case.  In fact it can be shown  that for any $d > 1$,  $r_c \,  \sim \, \sqrt{N} \,  L_{4+d}$. This subtlety  has to do with the singularity  properties of the higher co-dimension Greens functions.
    %They however are unimportant for our investigation.  

  \section{Solitonic Branes} 

We shall now consider another example, in which the cutoff of species 
theory is again bounded by the bulk holographic scale.  We shall see, that 
by consistency of the underlying theory and brane's inner structure, the distance below which species 
can be considered as 4D elementary particles ($l_{species}$) is secretly constrained by the bulk holographic scale $L_{N(4+d)}$. 
 We then try to deform the theory by taking 
$l_{species}\, \rightarrow \, 0$ and keeping $L_{N(d+4)}$ fixed. 
Again, we shall discover that theory responds by creating a strong coupling scale ($L_{strong}$) in 4D theory which is automatically bounded 
by the bulk holographic scale, according to (\ref{strongbound}).   
Even in the gravity-decoupling limit,  4D theory modes appear with the strong coupling scale bounded by $L_{N(4+d)}^{-1}$. 
 
  We shall consider solitonic branes that are solutions of classical field equations of a high-dimensional theory.  Since we are working in asymptotically flat space-times,  we shall choose the 
  codimension-two branes, which are known to produce asymptotically-flat spaces. 
  We shall consider a solitonic brane produced by a 6D scalar field $\Phi$, charged under an 
$U(1)$ gauge symmetry.   The Lagrangian of interest is a trivial generalization  of the Abelian  Higgs
model to 6D: 
\begin{equation}
\label{6dhiggs}
|D_A \Phi|^2    \, - \, {1 \over 2} \, \lambda^2 (|\Phi|^2 \, - \, v^2 \, )^2 \, - \, {1 \over 4} F_{AB}^2 \, .
\end{equation}
Here $A,B$ are the six dimensional indexes, and 
the minimal coupling to 6D gravity is assumed. 
Notice, that both,  the Higgs coupling $\lambda$ as well as the gauge coupling ($g$),  have inverse-mass dimensionality. 
The Higgs field develops a vacuum expectation value at the scale $v$. The brane solution in question is a well-known  Nielsen-Olesen flux tube
\cite{vortex} lifted  to 6D. The solution,  in the cylindrical coordinates 
$\rho, \phi$, has the following asymptotic form.  For  $\rho \rightarrow \infty$,  $\Phi \, = \,v {\rm e}^{i\phi}$, 
which due to non-trivial  topology forces  $\Phi = 0$ at $\rho = 0$. The brane core supports the 
magnetic flux in the direction $F_{56}$. 
 The brane has two cores, the magnetic core (the region with localized magnetic flux)  and the Higgs 
core (the region with non-zero Higgs energy).  The widths of these cores are set by 
the inverse masses of the gauge ($m_{gauge}^{-1}  \, = \, (gv)^{-1} $) and the Higgs 
 ($m_{Higgs}^{-1} \, = \, (\lambda v)^{-1}$) particles  respectively. In the limit when $g \, = \, \lambda$, 
known as Bogomolnyi-Prasad-Sommerfield (BPS)  limit,  the two cores are of the same size.  As it is well known, in this limit, the gauge repulsion between the  two vortexes is exactly compensated by the Higgs attraction, and the vortexes are in a neutral equilibrium.  For simplicity,  we shall adopt this limit. 
The brane tension, energy per unit 4D volume, is given by  $T \sim v^2$
(or to be more precise, $T\, = \, 2\pi v^2$ in BPS limit), and is independent of the the gauge and Higgs couplings.   
 
  Outside of the core, the metric of the brane  is a generalization 
of Vilenkin's cosmic string metric \cite{vilenkin}, 
\begin{equation}
\label{stringmetric}
d^2s \, = \, - \,  dx_{\mu}^2 \, +   \, d\rho^2  \, + \,  (1 - {\delta \over 2\pi}) \rho^2 d\phi^2 \, , 
\end{equation}
where $\delta \, = \, 2\pi {T \over M_6^4}$ (where a  factor of $4$ has been absorbed in 
renaming $M_6$). We thus see that metric produced by brane  is an 
asymptotically-flat metric with the angular deficit set by $\delta$.  Critical value of the tension for which the brane over-closes the space (in fact, turns it into a  a cylinder) is $T \, = \,  M_6^4 $ corresponding to 
the angular deficit of $\delta \, = \,  2\pi$. 
 Each brane supports the two localized zero modes, $\pi^{a}$ ($a = 1,2$), which are the Nambu-Goldstone bosons of  the spontaneously broken translation invariance in extra directions.  Their localization width is 
 set by the brane core thickness 
$l_{core} \, \equiv  \, (\lambda v)^{-1}$. 
  
   Consider now  $N$  coincident branes.  In the BPS limit, such a system 
 continues to support $2N$ massless  Goldstone bosons.   
   Since the tensions add up, the  deficit angle produced by such a system is 
   $\delta_N \,  = \, 2\pi  \, N \, {T \over M_6^4}$.  Defining the scale 
   $\Lambda_{str} \, \equiv \, L_{str}^{-1}\, \equiv \, T^{{1\over 4}}$ and requiring that the deficit is less than $2\pi$, we get that 
   the bound on the tension scale, 
  \begin{equation}
  \Lambda^{-1}_{str}   \,  \geqslant  \,  {N^{{1\over 4}} \over M_6} \, .
   \label{boundlambda} 
   \end{equation} 
 Thus, for $N$ coincident branes, the bound on the brane tension scale (the  Higgs VEV) is set by the bulk  holographic  scale for  $N$ bits of the information, 
   \begin{equation}
   \Lambda^{-1}_{str} \, \geqslant \, L_{N6}   \, . 
   \label{vevbound} 
   \end{equation}   
   
  The localization width of  the species $l_{species} \, = \, l_{core}\, = \, m_{Higgs}^{-1}$ automatically sets the cutoff on the species theory for the 4D brane observer, since for shorter distances species are no longer point-like.  Notice, that for the weakly coupled bulk theory
$l_{core} \, > \, L_{strong}$, since the Higgs and gauge 
particles are lighter then the soliton mass scale.  
Thus, for  the weakly coupled bulk theory,  $l_{species} \, > \, \Lambda^{-1}_{str}$, and the holographic bound is automatically satisfied. 
  
  Remarkably, the bulk holographic scale continues to bound the strong coupling scale of the 4D theory even in the limit in which  the particle localization width goes to zero.  
   Indeed,  let us  take the limit  $g, \lambda \, \rightarrow  \infty, ~~ N \rightarrow \infty,~~ 
   M_6 \, \rightarrow \, \infty $ while keeping $\delta$, $v$ ($\Lambda_{str}$) and  $L_{N6}$  fixed.   
   In this limit Higgs  and gauge bosons become infinitely heavy and decouple from the low energy theory.  Gravity decouples as well.  So the bulk low energy theory in this limit is trivial, but 
   the solitonic sector delivers massless modes.  
 Notice that in this limit the resolution length scale  $l_{species}$  for  4D species goes to zero. 
 So  naively (just as in 5D example) one may think that the surviving 4D theory has no memory about the 
 6D holographic bound.  However,  this is not the case.  
 The 4D theory gets strongly-coupled  at the brane tension scale.  The low energy fields are the $\pi$-fields with the Nambu-type  action, 
 \begin{equation}
 \int \, d^4x  \, \Lambda^4_{str} \,  \sqrt{{\rm det} \, \partial_{\mu}\pi^a\partial_{\nu}\pi^a} \, .  
 \end{equation}       
  Notice, that because of (\ref{boundlambda}), the strong coupling scale of the surviving 4D theory is 
  again set by the bulk holographic scale.  
  
   We thus  see, that whenever we try to decouple the species resolution scale,  $l_{species}$,  from 
   $L_{N6}$, the 4D theory responds by creating a sector with the  strong coupling scale 
   bounded by $\Lambda_{str} \, = \,  L_{N6}^{-1}$.

\section{Species on D-Branes}

 We shall now consider a case  when $N$ 4D species are localized on $D_3$-branes embedded in 
an asymptotically flat 10D space. As we shall see, the same property  holds true in this case also. 
Namely, the strong coupling scale of the 4D theory is bounded by the 10D holographic scale,  $L_{N10}$.   This follows from AdS/CFT correspondence.

  Indeed, in order  to get  $N$ 4D species embedded in  10D asymptotically-flat space, we   
start by considering a stuck  of  $N_D \,  \equiv \, \sqrt{N}$  $D_3$-branes.  This is exactly the setup of  AdS/CFT correspondence.    
 % In the case of the AdS/CFT correspondence we start with 10 dimensional gravity with the scale $R$ determined by the gravitational radius of $N$ D-3 branes, namely
 Because of their BPS properties,  $D_3$-branes produce a static asymptotically-flat metric  with 
the 10D gravitational  radius given by,  
\begin{equation}
R=(gN_D)^{\frac{1}{4}}l_s \, \equiv \, \lambda^{\frac{1}{4}}l_s \, ,
\end{equation}
with $l_s$ denoting the string length.   In the last expression we have defined the t'Hooft coupling $\lambda$. 
The first hint towards the holographic connection is already given
by the fact that  $R$ coincides with the 10-dimensional holographic scale $L_{N10}$  for the storage of 
$N$ bits of information,   and thus is solving the bound 
%Notice that the holographic meaning of $R$ is to fix the ten dimensional scale for storage of $N^2$ degrees of freedom. 
%In fact $R$ is solving the bound
\begin{equation}
N \, = \, (RM_{10} )^8 \, , 
\end{equation}
with the ten dimensional Planck mass $M_{10} \, = \, \frac{1}{g^{\frac{1}{4}} l_s}$. 
%Following our previous discussion we can consider ten dimensional degrees of freedom with wave length smaller than $R$.
 The effective near-horizon gravity is  $AdS_{5} \otimes S^5$ with metric
\begin{equation}
\label{2}
ds^2=\frac{r^2}{R^2}dx^2 + \frac{R^2}{r^2}(dr^2 +r^2 d\Omega_{5}) \, .
\end{equation}
 
 We shall now repeat some usual steps for establishing AdS/CFT  in order to  show how the 
 connection between the 4D strong coupling scale and $L_{N10}$ emerges.  
Since in the case of the AdS/CFT correspondence we are interested in a duality between a gravity theory in ten dimensions and a non gravitational theory in four dimensions we should proceed to analyze the gravity decoupling limit of the holographic correspondence discussed above. Generically the gravity decoupling limit can be defined in two different ways. Namely as the limit of 
$l_s \, \rightarrow  \, 0$,  or instead as the limit $ g \, \rightarrow \, 0$ but keeping the string length 
$l_s$ finite. Let us first consider the decoupling of gravity in the limit $l_s \, = \, 0$. By introducing the holographic energy variable $U=\frac{r}{l^2}$ we transform the metric (\ref{2}) into 
\begin{equation}
ds^2= l^2 ( \frac{U^2}{\lambda^{\frac{1}{2}}}dx^2 +\lambda^{\frac{1}{2}} \frac{dU^2}{U^2} +\lambda^{\frac{1}{2}}d\Omega_{5}) \, .
\end{equation}
Nicely enough string theory is perfectly well defined in this background even in the limit $l_s\, =\, 0$, therefore we get a full fledged gravity theory in ten dimensions. However, in this limit gravity in the four dimensional theory is completely decoupled since $M_{10}\, = \, \infty$. Moreover the scale 
$R \, = \, L_{N10}$ goes to zero in this limit leading to the well known CFT in four dimensions with infinite cutoff.

  In order to trace the holographic meaning of the 4D strong coupling scale, we need a limit in which 
  $R$ is kept finite.  
 As already pointed out a different gravity decoupling limit could be defined by sending $g_s$ to zero,  but keeping the string length $l_s$ finite. 
 This gravity decoupling limit becomes very useful for our purposes if at the same time we send $N$ to $\infty$ with $gN$ (and thus $R$ and $L_{N10}$) finite. The holographic meaning of  this t'Hooft limit for decoupling of gravity is  that - even with vanishing gravity- we still get a non vanishing gravitational radius for the storage of the $N$ bits of information.
  In the other words, the memory about the bulk holography persists in 4D theory in form 
  of the strong coupling scale.
 The connection now is clear,  the same scale  $R$ sets both  the strong-coupling scale of 4D theory
 with $N$ species as well as the 10D holographic scale for $N$ information bits.

  Let us now briefly discuss the holographic correspondence in this particular limit. In terms of the holographic energy variable $U$ and for finite $l_s$ the condition of wave length smaller than $R$ induces the existence of a UV wall in the holographic direction at $U=\lambda^{\frac{1}{4}}\frac{1}{l}$. Notice that this UV wall goes to infinity in the limit $l_s=0$. But,  if we keep $l_s$ finite,  the scale $R$ in the dual four dimensional theory still survives although gravity is completely decoupled. Indeeed, this scale is just $\lambda^{\frac{1}{4}} l$ that,  as already pointed out,  is the gravitational radius for the storage of $N^2$ bits localized in four dimensions. In summary,  when we introduce an  UV wall in the holographic direction at $U=\lambda^{\frac{1}{4}}\frac{1}{l}$ the ten dimensional gravity theory becomes dual to a four dimensional theory with gravity completely decoupled, but with a dynamical scale $R\, = \, \lambda^{\frac{1}{4}} l$ set by the bulk holographic  scale.   This is of course a sort of a "QCD" like 
 theory with the QCD scale given by $\Lambda_{QCD} \, = \, R^{-1}$. This "holographic" confinement is simply identifying the size of the singlet glueball with its gravitational radius, that , in the t'Hooft limit, is non vanishing even if gravity is completely decoupled.

\section{Duality and Localization: Attempting a Synthesis}

Generically holographic duality maps a higher dimensional bulk gravity theory into a lower dimensional theory on the boundary \cite{tHooft}. For the AdS/CFT correspondence the map is between type IIB string bulk theory on $AdS_{5} \otimes S^{5}$ with length scale $R$ and $N=4$ four dimensional SSYM with gauge group $SU(N_D)$ and t'Hooft coupling $\lambda$,  where $N_D^{\frac{1}{4}} = \frac{R}{l_{pl}}$ and $\lambda^{\frac{1}{4}} =\frac{R}{l_{st}}$. The bulk semiclassical gravity requires both $N$ and $\lambda$ to be large. From the ten dimensional point of view the holographic meaning of $N$ is to define a bound on the amount of information we can store in a ten dimensional space region of size $R$ \cite{SW}. 

 An alternative but equivalent interpretation of $N$ is the maximal number of different particle species compatible with Einstenian ten dimensional quasi-classical black holes of size bigger or equal to $R$.
Or equivalently, $N$ is a maximal number of 10D particle species compatible with the cutoff length  
$R$.  

  Therefore,  a general feature of any holographic duality is to relate a $4+d$  gravity theory character
 ized by a scale $R$ to a four dimensional field theory with $N$ species,  where $N$ is determined by the holographic relation $N^{\frac{1}{d+2}} =\frac{R}{L_{4+d}}$,  with $L_{4+d}$ being the Planck length in $4+d$ dimensions. This general feature of holographic dualities leads us to focus our attention on the dynamical mechanism of localization of particle species on a lower dimensional topological defect. Very likely we could be able to unravel from the dynamics of this localization mechanism the basic elements of the holographic correspondence between physical theories in different dimensions.

As we have described in this note the first consequence of localization of species on a lower four dimensional manifold is to set the strong coupling scale of the low energy 4D theory by  bulk 
holographic scale.  The strong coupling comes either from the fact that species can no longer be considered as elementary  above that scale, or from the modification of gravitational dynamics 
at the crossover scale $r_{c}\, \ll \, R$ where gravity actually changes from $4$ into $4+d$. In the simplest case of $d=1$ the dynamical origin of this crossover scale is physically quite transparent and can be 
uniquely traced to the existence of  4D Eintein-Hilbert term. 
 The required magnitude of this term is exactly compatible with the one-loop contribution of localized species to the graviton propagator.  However,  our conclusions hold regardless of its precise origin.  
Independently of the actual mechanism of generation of the four dimensional IR scale $r_{c}$ this scale is holographically bounded by the higher dimensional scale $R \, = \, L_{N(4+d)}$:
\begin{equation}
r_{c}  \, \geqslant \,  R
\end{equation}
The origin of this bound lies in  the fact that the number of species is precisely defined by the value of $R$ in $4+d$ Planck units. The modified gravity on the lower dimensional theory necessarily involves the propagation of extra modes that become strongly coupled at a certain scale $r_{*}$ \cite{strong, stronggia}.  This strong coupling scale is determined by the trilinear coupling of the extra scalar helicity of effective 4D graviton.

For  making the parallel with string theory case, let us generalize the notion of $r_*$ and $L_{str}$ distance, by 
defining the length scale at which
this trilinear coupling will be order $g^{2}$, 
\begin{equation}
\label{one}
L_{str} \, = \,  (r_{c}^2 L_4)^{\frac{1}{3}} g^{-\frac{2}{3}} \, ,
\end{equation}
with the threshold of strong coupling at $g=1$ \footnote{The $L_{4}$ in equation (\ref{one}) is the Planck length in four dimensions}. Using now the value of the  scale $r_{c}=\frac{N}{M_5}$ we easily get $L_{str} (N)  = \frac{N^{\frac{2}{3}}}{M_5} g^{-\frac{2}{3}}$. Now it is easy to observe that the strong coupling scale is playing a role  similar  to the string scale in the higher dimensional theory. Indeed $\frac{r_{c}}{L_{str}(N)} = N^{\frac{1}{3}}g^{-\frac{2}{3}} = \lambda^{\frac{1}{3}}$ for $\lambda =g^2N$ \footnote{Notice that if we identify the parameter $g$ with the string coupling constant and the number of species with $N_{g}^2$ for $N_{g}$ the number of colors $\lambda =g^2N$ is just the square of the standard t'Hooft coupling and the relation $\frac{r_{c}}{L_{str}(N)} = \lambda^{\frac{1}{3}}$ is just the expected AdS/CFT correspondence in five dimensions, namely $R=\lambda_{g}^{\frac{2}{3}}l_{s}$ for $\lambda_{g}= N_{g}g$ the standard t'Hooft coupling for a gaug
 e theory
  with $N_{g}$ colors.}. Moreover the semiclassical bulk gravity regime defined by $\lambda>1$ corresponds to the regime $r_{c} > L_{str}(N)$.

In summary,  it appears that there could be defined a general holographic analogy
 between a $(4+d)$-dimensional  string theory with scale $R$, on one side, and four dimensional modified gravity with $N$ localized species, on the other side, where the number of species is determined by $R$ in Planck units and where the scales $r_{c}$  and the strong coupling scale $L_{str}(N)$ are in one to one correspondence with the string theoretic  scales $R$ and $l_{s}$.

Notice that the above  discussion is completely general and only uses the fact that the higher dimensional gravity theory is characterized by the scale $R\, =\, L_{N(4+d)}$ with the holographic meaning of fixing the number of species in the lower dimensional theory.
% An important caveat is , of course, how much our general argument depends on the physical nature of the scale $R$ as defining a negative or positive cosmological constant.

   \section{Summary and Outlook} 
   
    In this work we have identified the holographic origin of the strong coupling scale in 
  4D theories with $N$ particle species that originate as world-volume theories on 
  4D surfaces (3-branes)  embedded in asymptotically-flat $(4+d)$-dimensional spaces with Einsteinian 
  gravity in the bulk.   
 
  We have seen, that by consistency of the theory, either the species-theory  or  the gravity of the 4D  sources (or both) necessarily require an UV completion above the bulk holographic scale  $L_{N(4+d)}$.  In this way, bulk  holography dictates the laws of 4D physics. 
  
  In some examples reconciliation with the bulk holography happens via species becoming effectively composite (non-point-like) at distances $l_{species}$ shorter than $L_{N(4+d)}$.  
   
  However,  even when the species compositeness scale is pushed all the way 
   down to the fundamental Planck length $L_{4+d}$,  the strong coupling is not avoided. 
  In such a case,  the 
    gravitational sector responds by crossover modification of Newtonian 
 force law  from $4 + d$- to $4$- dimensional regime at  some distance  $r_c \, \gg \, L_{N(4+d)}$. 
 This automatically implies that  the scalar polarization of graviton becomes  strongly coupled 
 at distance $r_*$,  determined precisely by the bulk holographic scale. In fact , the gravitational 
 field of a single elementary information bit becomes strong exactly at the scale $r_* \, = \, L_{N(4+d)}$. 
  
   In all the considered examples there is a consistent gravity-decoupling limit in which the system is reduced to a pure 4D theory with the strong coupling scale determined by the holographic scale of the high-dimensional gravity theory. 
   
     We shall now  briefly summarize possible implications of our findings.

  {\bf 1.  Towards Generalizations of AdS/CFT  Correspondence.}

    Our  observation,   that  $4+d$-dimensional  holographic scale  
  $L_{N(4+d)}$ sets the bound  on the  strong coupling scale of  the  4D theory,  partially relies on  the 
  asymptotic  flatness of the embedding space.  To be more precise,  
  the curvature radius of the high-dimensional space must exceed the holographic scale
  $L_{5(4+d)}$.

    For $D$-branes the feature of asymptotic flatness rests on their BPS properties.  The same property 
    guarantees that the near-horizon limit is AdS geometry.  As we have shown, the latter fact  automatically makes the holographic origin of the 4D strong coupling  understandable in terms of  AdS/CFT correspondence phenomenon. 
    
     On the other hand, we have seen that 4D strong coupling exhibits exactly the same holographic 
  origin in the examples  in which the asymptotic flatness  of the embedding space  does not necessarily 
   imply any AdS type near-horizon geometry.   For example, neither for the species on the tensionless orbifold planes, nor  for  the 6D vortexes this property was there.  Yet, the fact that the 4D strong coupling scale $L_{strong}$ was bounded by the bulk holographic scale $L_{N(4+d)}$ persisted 
  also in  these examples  in a rather profound way.   
    
   It thus emerges,  that that the high-dimensional holographic origin of  4D strong coupling 
 is a result of species number and asymptotic flatness of the brane metric, rather than necessarily of the AdS-type properties of the near-horizon geometry, or even of the BPS properties.  Although for the $D$-branes the latter properties happen to follow from  the former (and vice-versa),  in general this is not the case. 

    These findings indicate that the relation 
  $L_{N(d+4)}  \leqslant L^{(strong)}_{(4)}$  may be a  way to extend certain properties of 
  AdS/CFT duality to other spaces. Of course,  in each particular case, the extended duality  
  may not be as powerful as it is in the  AdS limit,  however it can still be extremely useful in understanding  certain phenomena  (such as the appearance of the strong coupling considered in this work) in terms of gravity duals.  
  
  {\bf 2.  In Search for Gravity-Duals.}
  
     In general, search for  gravity-duals of strongly coupled gauge systems may be an extremely complicated task. Our findings provide evidence that the strong coupling scale is linked to the holographic scale of the gravity-duals.  A toy explicit  model for exploring such a connection may be a generalization 
  of our vortex example to the non-Abelian string case with $CP_n$ world volume theory\cite{cpn}.
  An indication for the gravitational origin of the gauge theories on the branes, comes from the general fact that localization of the perturbatively-massless  gauge fields on the branes requires  condensation 
  of magnetic charges in the bulk, which automatically implies the existence of both open and closed 
  strings in such a theory\cite{giamisha}.  The  strings in question are the electric flux tubes  (or glueballs) whose existence follows from the gauge invariance and the charge conservation on the brane.  
   In this respect,  any brane with localized gauge fields shares similarity with $D$-branes, and thus 
 secretly knows about the bulk gravity. 
  
  {\bf 3.  Implications for Large Distance Modification of Gravity.} 
  
   One of the byproducts of our studies is that  any crossover modification of classical gravity at distance 
   $r_c$ can be understood in terms of holography,  even without any reference to particle species. 
   Indeed, such a modification automatically follows from the requirement that  a short-distance observer be able to store 
   $N$ information bits in the box of the size of {\it far-infrared } Planck length.   In our 5D example 
  of \cite{dgp} the far-infrared Planck length is the  5D Planck length $L_5$.  
  This requirement fixes the crossover scale to be  $r_c \, = \, N L_5$.   Holographic 
  origin of the 4D strong coupling  is then encoded in $r_*$-phenomenon.  
  Indeed, as we have seen, the   $r_*$-radius of a single  elementary information bit is always given by $L_{N5}$.  
   
    It has been shown \cite{stronggia},  that { \it any}  crossover modification of 4D Newtonian  gravity beyond some IR scale $r_c$, 
results in the strong coupling and  $r_*$-phenomenon at intermediate distances  $\ll \, r_c$ . 
This conclusion follows from  the general covariance and the fact that gravitons in such theories always contain extra scalar helicities by condition of  absence of ghosts.   
  Our suggestion about the holographic nature of the strong coupling in such theories
 provides evidence of their high-dimensional origin.   This evidence is also supported by the known fact that  all sensible theories of IR-modified gravity contain continuum (or finely spaced) tower  of spin-2  (and possibly spin-0)  states. This fact on 
 its own serves as  indication of  the high-dimensional origin of such theories in which the spin-2 tower is identified with KK states.  In combination with the strong coupling holography, 
 the evidence for high-dimensional origin becomes even stronger.   
 
  Another interesting question is a generalization of our results for theories with more than one crossover scale $r_c$. 
  In this respect we note, that some progress has been made recently \cite{cascade} in identifying  class of theories with multiple stages of crossover modifications  of gravity within the framework of so-called ``cascading'' generalizations of \cite{dgp}.    The idea is to have a sequence of branes of decreasing  dimensionality embedded within each other in such a way,  that  each subsequent brane changes the dimensionality  of Newton's law by one.  This framework can be a convenient  laboratory for studying  the holographic origin  of the 4D strong coupling in the presence of several crossover distances, and providing holographic dictionary in terms of number of species propagating in various dimensionalities.

   Finally, the  holographic-equivalence of  the large distance modified gravity  theories to the theories with $N$ species suggests that 
certain gravitational properties can be understood in terms of species dynamics.  
 Some equivalence between the  cosmological dynamics on the two sides was already demonstrated in \cite{barvin}.

  \vspace{5mm}
\centerline{\bf Acknowledgments}

The work of G.D is supported in part
by European Commission  under 
the ERC advanced grant 226371,  by  David and Lucile  Packard Foundation Fellowship for  Science and Engineering and  by the NSF grant PHY-0758032. 
The work of C.G. has been partially supported by the Spanish
DGI contract FPA2003-02877 and the CAM grant HEPHACOS
P-ESP-00346.


\begin{thebibliography}{99}


\bibitem{ADS}

 
J. M.~ Maldacena, 
 `` The Large N limit of superconformal field theories and supergravity", 
 Adv.Theor.Math.Phys.  {\bf 2 } (1998) 231, Int.J.Theor.Phys. {\bf 38}  (1999) 1113,  hep-th/9711200; 
 
  
S.S.~Gubser, I.R.~ Klebanov, A. M.~Polyakov,    
 ``Gauge theory correlators from noncritical string theory",  Phys. Lett. {\bf B428} (1998) 105, 
  hep-th/9802109;

 E.~ Witten,     ``Anti-de Sitter space and holography", Adv.Theor.Math.Phys. {\bf 2} (1998) 253,
 hep-th/9802150

\bibitem{tHooft} G.~'t Hooft,
"Dimensional reduction in quantum gravity", gr-qc/9310026;
L.~Susskind,
"The World As A Hologram", J.\ Math.\ Phys.\  {\bf 36}, 6377 (1995),  hep-th/9409089.



\bibitem{bound}

G.~Dvali, ``Black Holes and Large N Species Solution to the
Hierarchy Problem,'' arXiv:0706.2050 [hep-th];

G.~Dvali and M.~Redi, ``Black Hole Bound on the Number of Species and Quantum Gravity at LHC,''
Phys. Rev.  {\bf D77} ( 2008) 045027,  arXiv:0710.4344 [hep-th];

G.~Dvali and D.~Lust, ``Power of Black Hole Physics: Seeing through the Vacuum Landscape,''
JHEP, {\bf 06}  (2008) 47, ArXiv:0801.1287 [hep-th];

G.~Dvali, ``Nature of Microscopic Black Holes and Gravity in Theories with Particle Species'',  arXiv:0806.3801 [hep-th];`

G.~Dvali, M.~Redi, 
 ``Phenomenology of $10^{32}$ Dark Sectors", arXiv:0905.1709 [hep-ph]


R.~Brustein, G.~Dvali and G.~Veneziano, `` A Bound on Effective 
Gravitational Coupling from Semiclassical Black Holes", in preparation. 


\bibitem{us} G.~Dvali and C~Gomez, ``Quantum Information and Gravity Cutoff in Theories with Species'',
Phys. Lett. {\bf B674}  (2009) 303,  arXiv:0812.1940 [hep-th] .
 

     
\bibitem{loops1}
G.~Dvali and G.~Gabadadze, ``Gravity on a brane in infinite-volume extra space,''
{\it Phys. Rev.}  {\bf D 63}, 065007 (2001) [hep-th/0008054].

\bibitem{loops2} 
G.~Veneziano, ``Large-N bounds on, and compositeness limit of, gauge and gravitational
 interactions,''
  JHEP {\bf 0206}, 051 (2002)
  [arXiv:hep-th/0110129].

 \bibitem{Br1}
  R.~Brustein, D.~Eichler, S.~Foffa and D.~H.~Oaknin,
  ``The shortest scale of quantum field theory,''
  Phys.\ Rev.\  D {\bf 65}, 105013 (2002)
  [arXiv:hep-th/0009063].
  
  
\bibitem{Bek1}
  J.~D.~Bekenstein,
  ``Is The Cosmological Singularity Thermodynamically Possible?,''
  Int.\ J.\ Theor.\ Phys.\  {\bf 28}, 967 (1989).
  


\bibitem{Br2}
  R.~Brustein, D.~Eichler and S.~Foffa,
  ``A braneworld puzzle about entropy bounds and a maximal temperature,''
  Phys.\ Rev.\  D {\bf 71}, 124015 (2005)
  [arXiv:hep-th/0404230].

  R.~Brustein,
  ``The generalized second law of thermodynamics in cosmology,''
  Phys.\ Rev.\ Lett.\  {\bf 84}, 2072 (2000)
  [arXiv:gr-qc/9904061].

 
 \bibitem{dgp}
  G.~Dvali, G.~Gabadadze and M.~Porrati,
  ``4D gravity on a brane in 5D Minkowski space,''
  Phys.\ Lett.\ B {\bf 485}, 208 (2000); {\tt hep-th/0005016}.

\bibitem{vDVZ}
H.~van Dam and M.~J.~G.~Veltman,
 ``Massive And Massless Yang-Mills And Gravitational Fields,''
Nucl.\ Phys.\ B {\bf 22}, 397 (1970);
V.~I.~Zakharov,
``Linearized gravitation theory and the graviton mass,''
JETP Lett.\  {\bf 12}, 312 (1970);
 
     
\bibitem{loops3}
S.~Corley, D~.A~.Lowe and S.~Ramgoolam, JHEP {\bf 0107}, 030 (2001) [hep-th/0106067].

%\bibitem{strings2}

%I.~Antoniadis, R.~Minasian and P.~Vanhove, Nucl. Phys.  {\bf B648}, 69 (2003), [hep-th/0209030].

\bibitem{strong}
C.~ Deffayet, G.~Dvali, G.~Gabadadze and  A.I.~Vainshtein,  
``Nonperturbative continuity in graviton mass versus perturbative discontinuity", 
 Phys. Rev. {\bf D65}  (2002) 044026,  hep-th/0106001. 

 \bibitem{prl}
M.~ A.~ Luty, M.~Porrati and R.~Rattazzi, 
JHEP {\bf 0309}, 029 (2003) [hep-th/0303116].


\bibitem{stronggia} 
 G.~ Dvali,  
`` Predictive Power of Strong Coupling in Theories with Large Distance Modified Gravity",
 New J.Phys.  {\bf 8}  (2006) 326,  hep-th/0610013

 

\bibitem{lue} A.~Lue,
``Cosmic strings in a brane world theory with metastable gravitons",
Phys. Rev. {\bf D 66}, 043509 (2002), [hep-th/0111168].

\bibitem{andrei} A.~Gruzinov,
`` On the graviton mass", astro-ph/0112246.

\bibitem{moon} 
G.~Dvali, A.~Gruzinov, M.~Zaldarriaga, 
``The Accelerated universe and the moon",  Phys. Rev. {\bf D68} (2003) 024012,  hep-ph/0212069


\bibitem{ls} A.~Lue and G.~Starkman, Phys. Rev. {\bf D 67}, 064002 (2003) [astro-ph/0212083].


\bibitem{nicol} 
 Classical and quantum consistency of the DGP model.
A.~Nicolis, R.~Rattazzi,  JHEP {\bf 0406} (2004) 059,  hep-th/0404159. 


\bibitem{greg}
G.~Gabadadze and A.~Iglesias, 
Phys. Rev. {\bf D72}  (2005) 084024, [hep-th/ 0407049];
Phys. Lett. {\bf B 632} (2006) 622, [hep-th/0508201].

\bibitem{oriol}

G.~Dvali, G.~Gabadadze, O.~ Pujolas, R.~ Rahman, 
 ``Domain Walls As Probes Of Gravity",
  Phys. Rev. {\bf D75}  (2007) 124013,  hep-th/0612016. 


\bibitem{arkady}

 A.I.~Vainshtein,
 ``To the problem of nonvanishing gravitation mass,''
Phys.\ Lett.\ B {\bf 39}, 393 (1972).


\bibitem{BD}  D.G.~Boulware and S.~Deser, Phys.Rev. D {\bf 6} (1972) 3368; 


\bibitem{ghostnicol}

P.~Creminelli, A.~ Nicolis, M.~Papucci, 
 ``Ghosts in massive gravity",  JHEP {\bf 0509} 2005 003, hep-th/0505147

\bibitem{ghostcedric} C.~Deffayet and J.W.~Rombouts, Phys. Rev. D {\bf 72} (2005) 044003, 
gr-qc/0505134.


\bibitem{degrav}

 G.~Dvali, S.~Hofmann, J.~ Khoury, 
 ``Degravitation of the cosmological constant and graviton width'', 
Phys. Rev. {\bf D76}  (2007) 084006,  hep-th/0703027. 

\bibitem{greg2} 
G.~Gabadadze, A.~Iglesias, 
 ``(De)coupling limit of DGP",  Phys. Lett. {\bf B639}  (2006) 88,  hep-th/0603199. 


\bibitem{vortex}

H. B.~Nielsen and  P.~Olesen, 
 ``Vortex Line Models for Dual Strings", 
  Nucl. Phys. {\bf B61} (1973) 45. 


\bibitem{vilenkin}  A.~Vilenkin, Phys. Rev. {\bf D 23} (1981) 852. 





\bibitem{SW}

L.~ Susskind and  E.~ Witten, ``The Holographic bound in anti-de Sitter space", 
 arXiv:9805114 [hep-th]  

\bibitem{cpn}

D.~Tong, Quantum Vortex Strings: A Review. Annals Phys.324:30-52,2009. ( and references therein)

\bibitem{giamisha}

An incomplete list of references includes: 

G.~ Dvali, M. A.~Shifman, 
``Domain walls in strongly coupled theories", 
Phys. Lett. {\bf B396} (1997), Erratum-ibid.  {\bf B407}  (1997) 452,  hep-th/9612128. 

 A possible connection with large $N$ QCD string theory D-branes was first noticed by,   
E.~Witten,   ``Branes and the dynamics of QCD",  Nucl. Phys. {\bf B507} (1997) 658;   hep-th/9706109. 

G.~ Dvali, Z.~ Kakushadze 
`` Large N domain walls as D-branes for N=1 QCD string",
 Nucl. Phys. {\bf B537}  (1999) 297,   hep-th/9807140; 

G.~Dvali, G.~Gabadadze  and Z.~ Kakushadze, 
``BPS domain walls in large N supersymmetric QCD", 
  Nucl. Phys. {\bf B562} (1999) 158,  hep-th/9901032;  

G.~ Dvali, A.~Vilenkin, 
``Solitonic D-branes and brane annihilation", 
  Phys.Rev. { \bf D67}  (2003) 046002,  hep-th/0209217; 

M.~ Shifman and  A.~ Yung, 
 ``Domain walls and flux tubes in N=2 SQCD: D-brane prototypes".
 Phys. Rev. {\bf D67}  (2003) 125007,  hep-th/0212293. 

  
D.~ Tong, 
 ``D-branes in field theory",  JHEP 0602:030,2006,  hep-th/0512192. 

 \bibitem{cascade} 
    
C.~ de Rham, G.~Dvali, S.~Hofmann, J.~Khoury, O.~Pujolas, M.~Redi, A.J.~ Tolley, 
 ``Cascading gravity: Extending the DGP model to higher dimension'',
  Phys. Rev. Lett.  {\bf 100} (2008)  251603,  arXiv:0711.2072 [hep-th];

C.~ de Rham, S.~Hofmann, J.~Khoury, A.J.~ Tolley,  
``Cascading Gravity and Degravitation'', 
JCAP {\it 0802} (2008)  011,  arXiv:0712.2821 [hep-th];
    
O.~ Corradini, K.~ Koyama, G.~Tasinato,   ``Induced gravity on intersecting brane-worlds. Part I. Maximally symmetric solutions'',  Phys. Rev. {\bf D77} (2008) 084006, arXiv:0712.0385 [hep-th];
 ``Induced gravity on intersecting brane-worlds. Part II. Cosmology'',  Phys. Rev. {\bf D78}  (2008) 124002,  arXiv:0803.1850 [hep-th]

M.~Minamitsuji,  ``Self-accelerating solutions in cascading DGP braneworld'',   
arXiv:0806.2390 [gr-qc]
  
C.~ de Rham, J.~ Khoury, A.J.~ Tolley, 
``Flat 3-Brane with Tension in Cascading Gravity'',   
    arXiv:0907.0473 [hep-th]  

\bibitem{barvin}  


A.O.~ Barvinsky, C. ~Deffayet,  A.Yu. ~Kamenshchik,
Anomaly Driven Cosmology: Big Boost Scenario and AdS/CFT Correspondence.
JCAP 0805:020,2008,  arXiv:0801.2063 [hep-th]



  
 \end{thebibliography}
\end{document}